**Article**

# The effects of ecological selection on species diversity and trait distribution: predictions and an empirical test

Niv DeMalach*[1,2] (niv.demalach@mail.huji.ac.il)

Po-Ju Ke[1,3]

Tadashi Fukami[1]

[1]Department of Biology, Stanford University, Stanford, California, USA

[2]Institute of Plant Sciences and Genetics in Agriculture, Faculty of Agriculture, Food and Environment, The Hebrew University of Jerusalem, Rehovot, Israel

[3]Department of Ecology and Evolutionary Biology, Princeton University, Princeton, New Jersey, USA

* Corresponding author

## Abstract

Ecological selection is a major driver of community assembly. Selection is classified as stabilizing when species with intermediate trait values gain the highest reproductive success, whereas selection is considered directional when fitness is highest for species with extreme trait values. Previous studies have investigated the effects of different selection types on trait distribution, but the effects of selection on species diversity have remained unclear. Here, we propose a framework for inferring the type and strength of selection by studying species diversity and trait distribution together against null expectations. We use a simulation model to confirm our prediction that directional selection should lead to lower species diversity than stabilizing selection despite a similar effect on trait community-weighted variance. We apply the framework to a mesocosm system of annual plants to test whether differences in species diversity between two habitats that vary in productivity are related to differences in selection on seed mass. We show that, in both habitats, species diversity was lower than the null expectation, but that species diversity was lower in the more productive habitat. We attribute this difference to strong directional selection for large-seeded species in the productive habitat as indicated by trait community-weighted-mean being higher and community-weighted variance being lower than the null expectations. In the less productive habitat, we found that community-weighted variance was higher than expected by chance, suggesting that seed mass could be a driver of niche partitioning under such conditions. Altogether, our results suggest that viewing species diversity and trait distribution as interrelated patterns driven by the same process, ecological selection, is helpful in understanding community assembly.

**Keywords: annual plants, community weighted mean, community weighted variance, competition, environmental filtering, functional diversity, functional traits, seed mass, seed size, species richness, species pool, trait selection**

## INTRODUCTION

One of the major goals of ecology is to understand the mechanisms behind patterns of species diversity and trait distribution (Grime 1979, Tilman 1982, Huston 1994, Chesson 2000, Chase and Leibold 2003, HilleRisLambers et al. 2012). The theory of ecological communities (Vellend 2010, 2016) argues that variation among communities results from four high-level processes: ecological selection, dispersal, ecological drift, and speciation.

Ecological selection, the biotic and abiotic filtering of species from the species pool of potential colonizers, has been a major focus of studies on species diversity (Vellend 2016). Selection can be characterized by its strength, which increases with increasing differences in reproductive success among species (ecological fitness sensu Vellend 2016). Additionally, selection is characterized by type, it is stabilizing when species with intermediate trait values gain the highest reproductive success, whereas it is directional when species with extreme trait values gain the highest fitness (Fig. 1). Originally, this classification was proposed for the evolutionary selection of phenotypes within populations but it is currently applied for the ecological selection of traits within communities (Shipley 2010, Vellend 2016, Loranger et al. 2018) as we do here.

Both types of selection are thought to reduce community-weighted variance (CWV) of the traits being selected but have different effects on the community-weighted mean (CWM) (Rolhauser and Pucheta 2017, Loranger et al. 2018). The CWM should be similar to the mean of the trait distribution of the species pool under stabilizing selection, whereas it should vary under directional selection. Species diversity is expected to decrease with increasing selection strength (Vellend 2016) but the relationships between selection type and species diversity have not been fully investigated.

Since species diversity and trait distribution are both driven by ecological selection, we argue that they should be studied together within the same framework. We, therefore, propose to characterize selection attributes (i.e., strength and type) by simultaneously studying species diversity and trait distribution against null expectations built from the species pool. Specifically, let us assume the same species pool is shared across different communities, which could be different sites or habitat types within a heterogeneous landscape. Additionally, we assume that trait differences among species reflect competitive hierarchy instead of niche partitioning, i.e., ecological fitness decreases with increasing trait distance from the optimal value. These simplifying assumptions follow previous models of trait selection in ecological communities (Shipley 2010, Loranger et al. 2018) and their consequences are discussed toward the end of the manuscript.

Under the above assumptions, one would have evidence that selection had taken place if species diversity was lower than null expectation (Fig. 2A). Then, a hypothesis that a specific trait has been under selection would be supported if the observed CWV is lower than the null expectation (Fig 2B). We can also infer whether the selection was stabilizing or directional based on the deviation of the observed CWM compared to the null expectation (Fig. 2C). A more directional selection is expected to increase the deviation between observed CWM and the null expectation.

We investigated the proposed framework using a simulation model and apply it to an experimental case study. Our simulation model aims to verify the logic of the framework, i.e., that CWM and CWV can indicate the type and strength of ecological selection, respectively. Another aim of the model was to investigate the effects of different selection types on species diversity. Specifically, we expected that when trait distribution in the species pool is unimodal, directional selection would lead to lower species diversity than stabilizing selection because there are fewer species with extreme values.

To demonstrate the utility of our framework for understanding community assembly, we reanalyzed data from a mesocosm experiment of annual plant communities (Ron et al. 2018, DeMalach et al. 2019). This particular case study was chosen because it was designed as a selection experiment where the same set of species was sown under different levels of resource availability, thereby affecting the selection processes. Furthermore, for this system, we expected that one treatment would undergo stabilizing selection while the other would experience directional selection. Based on theoretical expectations (DeMalach and Kadmon 2018), we predicted that under low resource availability selection on seed size would be weak and stabilizing while under high resource availability it would be strong and directional because light competition favor large-seeded species. Together, the simulations and the case study show that the selection mechanisms can be identified only when trait distribution and species diversity are measured simultaneously.

**METHODS**

**Simulation model**

Our spatially implicit model describes population dynamics in a meta-community comprised of $n$ local communities. Competition occurs within each local community, and the local communities are connected by dispersal. For simplicity, the model assumes that the local communities have a fixed size and that there is no overlap among generations, as in annual species. In each time step, proportion $D$ of the community arrives from other local communities (hereafter dispersers), proportion $I$ arrives from outside the meta-community (hereafter immigrants), and the rest are descendants of individuals from the local community (hereafter residents).

Among each of the residents, the probability of belonging to species ($P_i$) in timestep $t + 1$ is determined by the following equation:

$$P_{i(t+1)} = \frac{\omega_i f_{i(t)}}{\sum_{j=1}^{S} \omega_j f_{j(t)}},$$  (1)

where $f_{i(t)}$ is the frequency of species $i$ in the local community (in the previous time step), $\omega_i$ is its ecological fitness (mean reproductive success), and $S$ is the number of species in the species pool (potential colonizers). A similar probabilistic rule applies for dispersers with the only difference being that meta-community frequency is used instead of the local community (i.e. they have an equal chance to arrive from all the local communities). All species have the same (extremely low) probability to arrive as immigrants from the species pool.

The ecological fitness of each species is determined by its specific trait value $\delta_i$ based on the following Lorentzian function:

$$\omega_i = \frac{1}{1+\theta(\delta_{best}-\delta_i)^2} \qquad , \qquad (2)$$

where $\delta_{best}$ represents the optimal trait value and $\theta$ determines the strength of the selection, i.e., the degree of the fitness differences for a given trait distance. When $\theta = 0$, all species have equal fitness while increasing $\theta$ intensifies fitness differences. This **Lorentzian function** was used to restrict $\omega_i$ to be positive for all values of selection strength. The trait values ($\delta_i$) in the species pool were assumed to be normally distributed ($\delta_{mean}$ and $\delta_{SD}$ are the mean and the SD of this distribution). For simplicity, the simulations focused on two scenarios, representing two extremes of a continuum. In the first, $\delta_{best}$ was equal to the mean value of the species pool (hereafter 'stabilizing selection'). In the second, it was equal to the highest value in the pool (hereafter 'directional selection').

All local communities started from a uniform abundance distribution of all species. We ran the model for 5000 timesteps. Visual inspection suggests that communities reached equilibrium by approximately 3000 timesteps, Appendix S1, Fig. S1-S3). We conducted three simulation runs for each parameter combination that we investigated (differences were minor). All the results represent the means of the three simulation runs, averaging from time step 3000 to 5000. The description of parameters and their values in the simulations are found in Table 1. In appendix 1, we tested the robustness of the model by investigating different assumptions about the trait distribution of the species pool (Appendix S1, Fig. S4-S6), alternative fitness function (Appendix S1, Fig. S7-S9), and a scenario when the trait being selected is not the trait being measured (Appendix S1, Fig. S10).

For each simulation, we calculated CWM and CWV. Additionally, we calculated species diversity under two scales (local communities and metacommunity), in terms of species richness and inverse Simpson index (hereafter Simpson diversity).

**Mesocosm experiment**

We applied the framework to a mesocosm experiment of annual plants growing in two habitats varying in soil depth (55 cm and 18 cm) and therefore productivity (Ron et al. 2018, DeMalach et al. 2019), hereafter referred to as the productive and the less productive habitat, respectively. We focused on seed mass selection patterns because previous analysis (DeMalach et al. 2019) has shown that it is the main predictor for abundance patterns along natural and experimental soil depth gradients (other measured traits were found to be insignificant).

A detailed description of the experimental system is found in the original papers (Ron et al. 2018, DeMalach et al. 2019). Briefly, the experiment was conducted at the botanical gardens of the Hebrew University of Jerusalem in Israel and consisted of nine artificial plant communities for each soil depth category. The mesocosm communities were established within metal containers with an area of $1 \times 1$ m. In December 2011, 51 annual species were sown in equal density (200 seeds per species, a total of $51 \times 200 = 10,200$ seeds per container). The species emerging in each container were let to grow and interact for five successive years (2011–2016) following their germination. All containers were blocked against dispersal (using mesh nets), which enables interpreting all patterns as consequences of selection and drift only.

At the (experimental) species pool level, seed mass was log-normally distributed (Appendix S2 S2, Fig. S1, Table S1) and therefore our analysis was based on $\log_{10}$-transformed seed mass data (as in most analyses of seed mass patterns). We used abundance data from the fifth growing season in a fixed quadrat of 25×25 cm at the center of each container for calculating species diversity (species richness and inverse Simpson's index) and seed mass patterns (CWM and CWV).

To test whether selections have occurred (regardless of which trait was selected for, Fig 2A) we compared the species diversity patterns with a simulation model of drift dynamics (i.e., $\theta = 0$) based on the specific parameter values of the mesocosm system (Appendix S1, Table S2). Although the experiment included 51 species, in the drift model species pool size was set to 47 based on the number of species blooming during the first year. This conservative assumption aimed to avoid naïve evidence for selection based on technical artifacts (e.g., non-viable seeds) of the experiment. The estimation of community size was based on the mean number of individuals measured in each container multiplied by 16 (the ratio between the sampled area and the total community). Initial composition was assumed to be a random sample from a multinomial distribution where all species have the same chance to be sampled (since sowing density was equal). To generate a distribution of outcomes we ran the drift model 1,000 times (for each iteration we calculated the mean of nine communities)

Based on our simulation results (Appendix S1, Fig. S10), a decrease in CWV could be driven by a selection acting on a different uncorrelated trait than the specific traits under investigation. While one could use dynamic simulations to determine whether the selection was specifically related to seed mass, this would require imposing assumptions from the theoretical model on the empirical data (e.g., the specific function relating fitness to traits). Instead, for CWM and CWV we used a trait-shuffling approach to generate the null expectation. We used the observed species abundance distribution from the mesocosm communities, assigned for each species a random seed mass value from the 'species pool' (i.e., the seed mass values of the sown species), and calculated the average of the nine communities in each treatment. Then, we compared the observed patterns of CWM and CWV to the expectation from 10,000 different randomizations.

## RESULTS

### Simulation

The model supports the assumptions of our framework that the CWM is mainly determined by selection type and CWV is mainly determined by selection strength (Fig. 3). The CWM differs from the mean of the species pool only under directional selection while the CWV is almost unaffected by selection type.

As expected, species diversity decreases with increasing selection strength (Fig. 3). However,

under any given level of selection strength, diversity is lower under directional selection. These

results are robust to the scale (local scale vs. meta-community scale) and the diversity indices

(Simpson diversity vs. species richness). Moreover, transient dynamics are qualitatively similar to equilibrium results (Appendix S1, Fig. S1-S3).

The lower diversity under directional selection is driven by two mechanisms operating in the same direction. First, as the species pool's trait distribution is normally distributed, species with intermediate traits have more similar fitness, which reduces extinction rate and enhances diversity under a given level of colonization. Accordingly, the difference between the two selection types is smaller under a uniform trait distribution in the species pool (Appendix S1,

Fig. S4-6). Still, diversity is higher under stabilizing selection, even under a uniform species pool

trait distribution because of boundary constraints (similar to the mid-domain effect; Letten et al.

2013), where only under stabilizing selection species with traits close to the optimum are found in both sides of the optimum. Our findings seem general and not restricted to the function we have chosen, because similar results were observed under alternative fitness function (Appendix S1, Fig. S7-9) and other alternative assumptions (Appendix S1, Fig. S1-S10).

**Mesocosm experiment**

Species diversity was lower in the productive habitat (Fig. 4). In both habitats, species diversity was lower than the null expectations generated from the drift model, indicating that selection has occurred. In accordance with our hypothesis, CWV was lower and CWM was higher than null expectations in the productive habitat, implying a directional selection for large-seeded species. In the less productive habitat, CWM was similar to the null expectation. In contrast with our hypothesis of stabilizing selection, CWV in the less productive habitat was higher (rather than lower) than the null expectations.

**DISCUSSION**

Recently, it was proposed that characterizing the type and strength of ecological selection could improve our understanding of the drivers of community assembly (Shipley 2010, Vellend 2016, Loranger et al. 2018). We, therefore, proposed a framework aiming to infer the underlying selection characteristics by measuring species diversity, CWM, and CWV in the local community and trait distribution in the species pool. We have demonstrated the utility of our framework by reanalyzing data from a mesocosm experiment that included two habitats that vary in their productivity. By comparing species diversity to null expectations, we showed that selection took place in both habitats but species loss was more severe in the productive habitat. The differences in species diversity probably arose from strong directional selection in seed mass that occurred only in the productive habitat. Below, we discuss our findings, elaborate on the simplifying assumptions of our approach, and highlight its implications.

**CWM and CWV as proxies for selection type and strength**

A major assumption of our framework is that CWM indicates the type of selection processes. The results of the model support this interpretation but CWM is only a proxy that should be interpreted with caution (see also Muscarella and Uriarte 2016). Under equilibrium, CWM can indicate the optimal strategy and therefore represents the degree of directionality in the trait selection. However, the rate at which equilibrium is reached is shorter with increasing selection strength (Appendix S1, Fig. S1-S2). During transient dynamics, deviation from the species pool is expected from the combined effect of strength and directionality.

CWV was found to be a reasonable proxy for selection strength because it is not strongly affected by selection type. Given that species diversity was higher under stabilizing selection when a given selection strength was applied, one might have expected communities under stabilizing selection to have higher CWV. In our model, there is a constant rate of immigration that prevents the species with the optimal strategy from monopolizing the community. Under stabilizing selection there are more species with trait values close to the optimum. These species have a long persistence time leading to higher species diversity but their contribution to CWV is minor because their trait values are close to CWM. In other words, the addition of species with traits close to CWM increases species diversity but its effect on CWV may not be positive. In some cases, there are some minor effects of selection type on CWV (Appendix S1, Fig. S4, S6, S9). We speculate that these differences are driven by different geometric constraints of stabilizing selection.

**Simplifying assumption of the simulation model**

Our model assumes that ecological fitness is affected by a single trait. However, our framework is not limited to this simplistic assumption. For multiple correlated traits that are difficult to disentangle, ordination techniques could be used as commonly done for sets of leaf traits in plant ecology ('the leaf economic spectrum', Diaz et al. 2016). Furthermore, our framework can be applied also for multiple uncorrelated traits by using a multidimensional trait space where directionality is characterized by the distance between trait values and the centroid of the species pool. Similarly, multidimensional trait dispersion indices can be used instead of community trait variance (Botta-Dukát 2005, Laliberté and Legendre 2010).

Our model also assumes that dispersal distance is equal for all species and therefore dispersal probabilities are only affected by reproductive output in the source population. The difference in dispersal distance among species can complicate the interpretation of our framework as it becomes difficult to disentangle differences in reproductive output and dispersal potential (Lowe and McPeek 2014). For sessile organisms like plants, a simple solution that was applied in the case study is blocking dispersal.

In the model, we focused on the simplest kind of selection, ('frequency-independent selection' sensu Vellend 2016), where trait differences among species affect only vital rates and\or competitive hierarchy.  Alternatively, trait selection can be frequency-dependent if traits are associated with niche partitioning and feedbacks (Vellend 2016). For simplicity and following previous models of trait selection (Shipley 2010, Loranger et al. 2018), we chose to focus only on frequency-independent selection as a starting point for integrating patterns of species and traits within the same framework. However, the results of the mesocosm experiment could not be explained solely based on frequency-independent selection (see the section below) which highlights the need to incorporate niche-partitioning into future extensions.

**Interpretation of the mesocosm experiment**

We sought to explain patterns of species diversity in the mesocosm experiment based on seed mass selection. A previous analysis of this system (DeMalach et al. 2019) has shown that seed mass is the main predictor of species habitat preferences which raised a hypothesis that species diversity patterns in that systems are related to seed size selection. Here, we tested this hypothesis by comparing the observed patterns of CWM, CWV, and species diversity to null expectations. Specifically, we predicted a strong directional selection under high productivity and a weaker stabilizing selection under low productivity. These predictions were driven by a resource competition model (DeMalach and Kadmon 2018) suggesting that large-seed species (that produce large seedlings) will be favored under high productivity where light competition is more intense.

In both habitats, selection has taken place, as indicated by species diversity being lower than the null expectation (Fig. 2). As we predicted, in the productive habitat CWM was higher and CWV was lower than null expectations, implying a strong and directional selection for large seed mass. This strong directional selection is probably one of the main drivers of the low diversity under these conditions.

In the less productive habitat, we found that in contrast with our prediction of stabilizing selection, CWV was *higher* (rather than lower) than the null expectation. If CWV was not different from the null expectation, the interpretation would be that traits other than seed mass were selected for (Fig. 2B). However, the finding that CWV was higher than null expectation suggests some kind of niche partitioning, where species with more distant trait values are more likely to coexist (limiting similarity sensu Macarthur and Levins 1967).

The maintenance of limiting similarity in seed mass is often explained by a trade-off between higher fecundity of small-seeded species and higher stress tolerance of large-seeded species (Muller-Landau 2010, D'Andrea and O'Dwyer 2021). This tradeoff enables the coexistence of species varying in seed mass when there is spatial heterogeneity in microhabitat quality and a positive correlation between seed mass and stress tolerance. However, in this system, spatial heterogeneity was minimized (Ron et al. 2018) and there was no evidence of higher stress tolerance of large-seeded species (DeMalach et al. 2019). We, therefore, attribute the coexistence of plants with different seed mass to a competition-fecundity tradeoff that enables small-seed species to grow in small microsites left unoccupied by the less fecund large-seeded species (Rees and Westoby 1997, Geritz et al. 1999, Coomes and Grubb 2003).

**Implications**

Our framework and model produce new predictions regarding the relationship between species diversity and functional diversity across environmental gradients (Mayfield et al. 2010, Cadotte et al. 2011, Rapacciuolo et al. 2019). When using CWV as the indicator for functional diversity, our model predicts that the two aspects of diversity should be positively correlated if diversity patterns along an environmental gradient are mainly determined by changes in selection strength (i.e., species diversity and CWV vary in the same direction with underlying selection strength). However, we expect a weaker correlation if diversity patterns along an environmental gradient are determined mostly by changes in selection type (e.g., a transition from stabilizing selection to directional selection affects only species diversity but not CWV; Fig. 3).

It may seem that there is a clear dichotomy between trait-based and species-based approaches (Shipley et al. 2016). However, explanations for ecological patterns often involve selection that affects both species diversity and trait distribution (e.g., Grime 1979, Tilman 1982). Our case study demonstrates that simultaneous investigation of species diversity and trait distribution helps to produce new insights into community assembly even in a system where both species diversity (Ron et al. 2018) and trait distribution (DeMalach et al. 2019) have been thoroughly investigated. We hope that our proposed approach will assist in shedding light on the underlying selection processes in many other communities.

## ACKNOWLEDGMENTS

We thank Ronen Ron and Ronen Kadmon for contributing data from their mesocosm experiment. Susan Harrison, Gili Greenbaum, and members of the community ecology group at Stanford University provided constructive comments on earlier drafts. This work was supported by the Rothschild fellowship (N.D.), Terman Fellowship of Stanford University (T.F.), and the studying abroad scholarship of the Ministry of Education of Taiwan (P.-J.K).

## AUTHORS CONTRIBUTIONS

N.D., P.-J.K, and T.F. conceived and designed the framework and analyses. N.D. performed the simulations and empirical analyses. N.D. wrote the first draft. All authors substantially contributed to the writing of the manuscript

## DATA ACCESSIBILITY

The empirical analysis was based on an available database:

https://figshare.com/s/5c18c0736b976df46e3c. The code of the simulations will be available upon publication.

**FIGURES & TABLES**

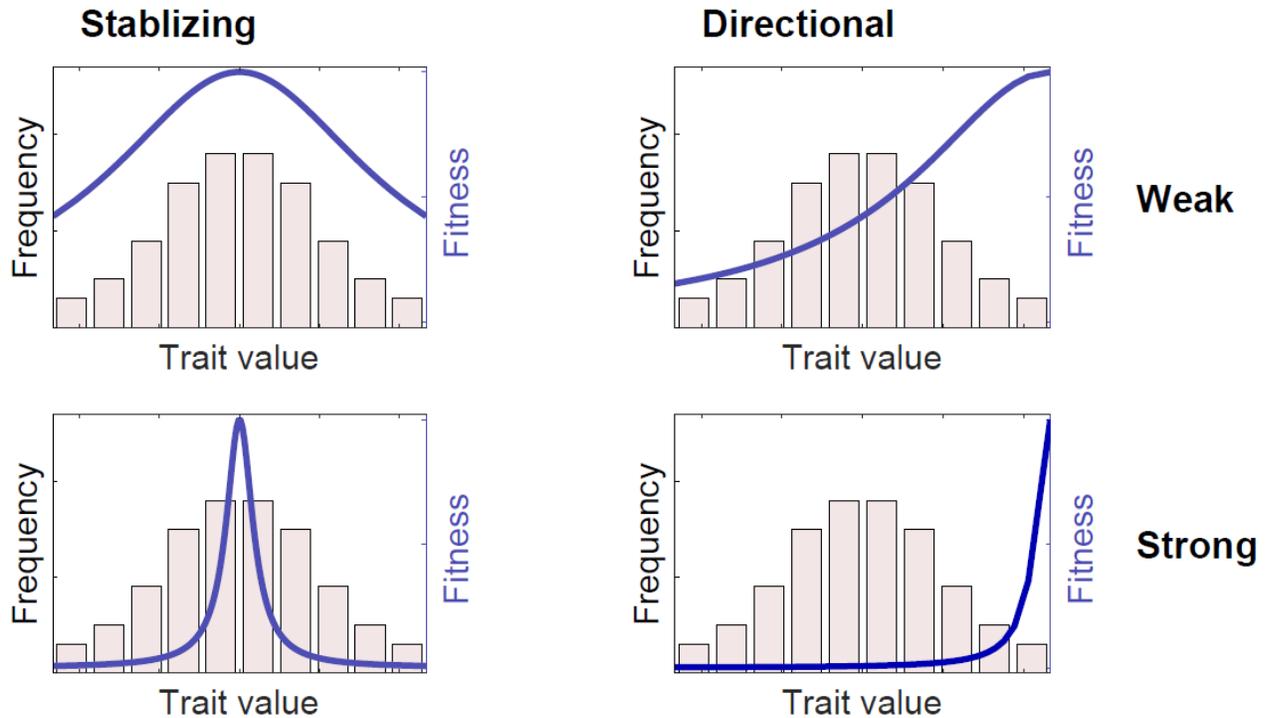

**Figure 1: Illustration of stabilizing and directional trait selection in ecological communities**. The blue curves represent the ecological fitness of different species (mean reproductive success) as a function of their trait values and the bars indicate species trait distribution in the species pool (one value per species). Stabilizing selection is when intermediate trait values of the trait distribution of the species pool matches the peak of the fitness curve, while directional selection is when the highest fitness is found under extreme trait values (in this example, for the highest values). Selection strength represents the degree of fitness reduction with increasing distance from the optimal value, where a steeper decline indicates stronger selection because of higher interspecific fitness differences.

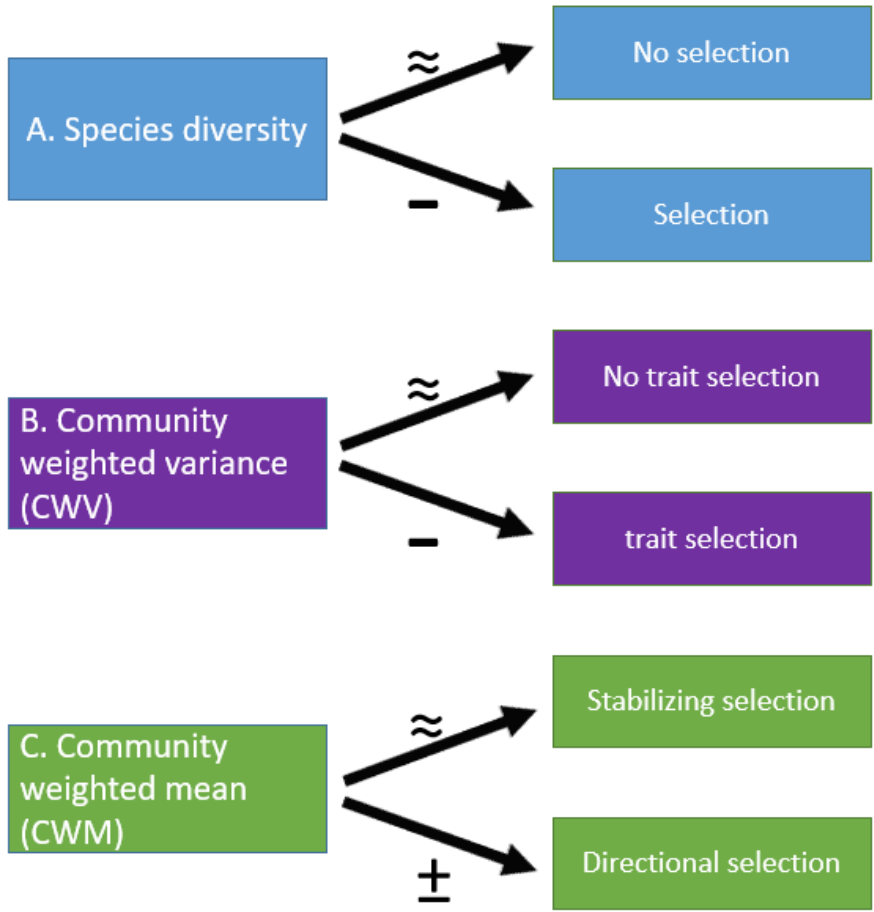

**Fig. 2**: A scheme for inferring selection attributes based on the comparison between observed patterns and null expectations. A) If species diversity is lower than the null expectation then selection has occurred (the larger the difference the stronger the selection). B) If community weighted variance (CWV) of a particular trait is lower than the null expectation it implies a trait-specific selection for that trait (the larger the difference the stronger the selection) C) If community weighted mean (CWM) of a particular trait differ from the null expectation it implies a directional selection. Alternatively, if CWM is similar, it implies a stabilizing selection (assuming the previous step has shown that CWV is lower than expected).

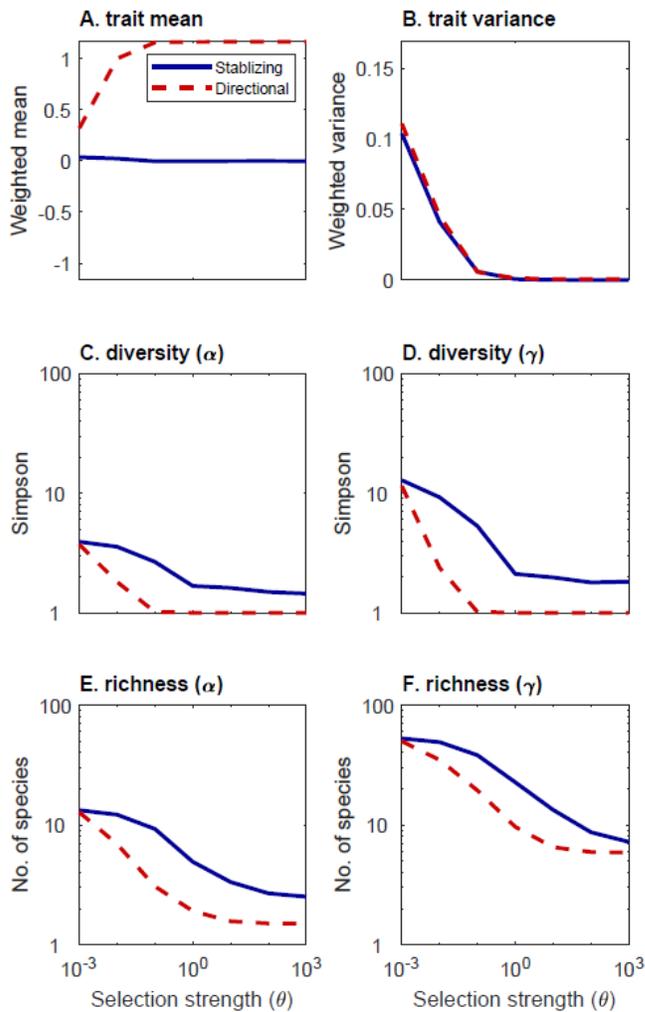

**Figure 3:** The community-level outcomes of varying selection attributes in the simulation model. (A) community weighted mean (CWM) of trait values diverge from the species pool mean (0) only under directional selection (B) community weighted variance (CWV) is affected by selection strength. Species diversity is affected by both directionality and selection strength: (C) Inverse Simpson diversity at the local community scale ($\alpha$); (D) Inverse Simpson diversity in the metacommunity scale ($\gamma$); (E) Species richness in the local community scale ($\alpha$); (F) Species richness in the metacommunity scale ($\gamma$). Note the logarithmic scale of the x-axes (all panels) and some y-axes (panels C-F).

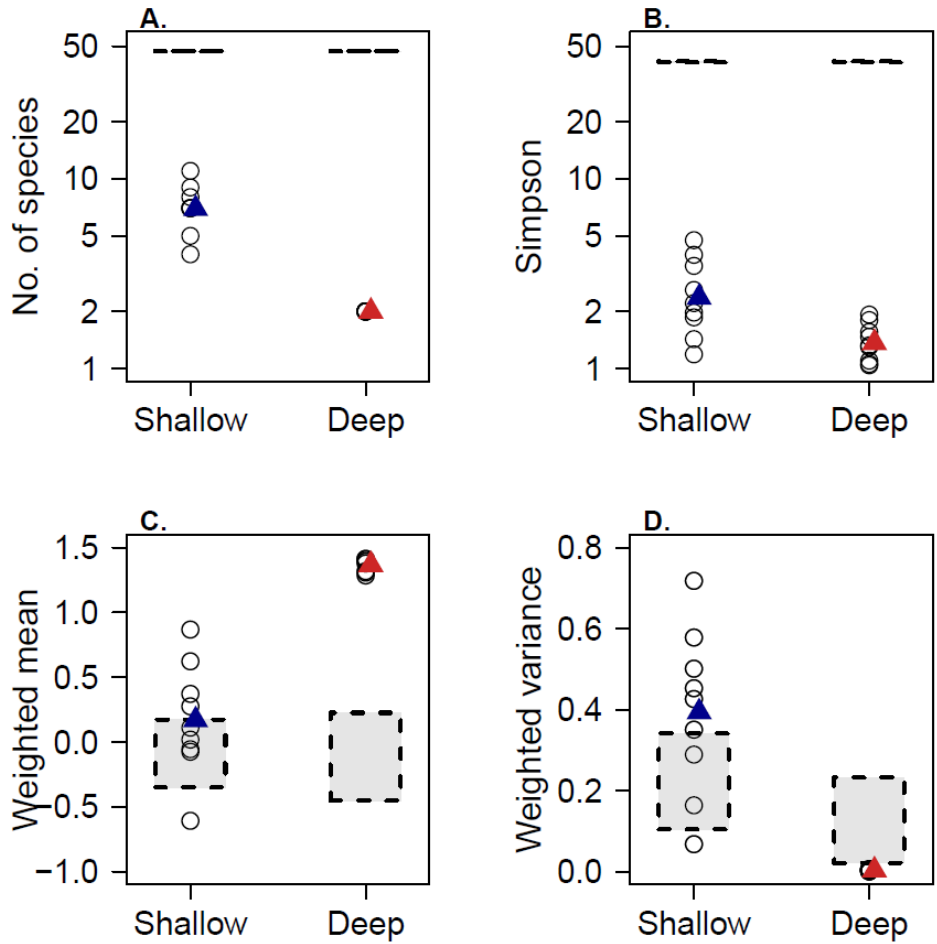

**Figure 4:** Ecological selection in the mesocosm experiment**.** Circles are values from each experimental community (n=18) while triangles represent the means of each soil depth treatment. Dashed lines represent the null expectations (a simulation envelope representing the extremum of 95% of the runs) (A) Species richness is lower in the deep soil treatment (productive habitat) compared with the shallow soil treatment (less productive) but lower than the null expectation in both habitat types. (B) Inverse Simpson's diversity is lower in the deep soil treatment, but in both habitat types, observed values are lower than the null expectations. (C) Community weighted mean seed mass [(log$_{10}$ mg)] is higher than the null expectation in the deep soil treatment. (D) Community weighted variance is higher than the null expectation in the shallow soil and lower than expected in the deep soil treatment. Note the logarithmic scale of the y-axes for diversity indices. Differences between the habitats are statistically significant (P<0.05 for all comparisons based on permutation tests using the R package 'Perm' for avoiding heteroscedasticity). See methods for details on the different null expectations for species diversity (based on a dynamics simulation of pure drift) and trait distributions (based on observed patterns of species distribution and the shuffling of trait values).

**Table 1. parameters of the simulation model**

| Symbol | Description | Value(s) |
|---|---|---|
| $S$ | Species pool size (number of species) | 100 |
| $N$ | Local community size (number of individuals) | 1000 |
| $n$ | Number of local communities | 10 |
| $\delta_{mean}$ | Trait mean (arbitrary units) | 0 |
| $\delta_{SD}$ | The standard deviation of the trait (arbitrary units) | 0.5 |
| $\theta$ | Selection strength (dimensionless) | $10^{-3}$-$10^{3}$ |
| $D$ | The proportion of dispersed individuals | $10^{-3}$ |
| $I$ | The proportion of immigrants from the species pool | $5 \cdot 10^{-4}$ |

# FIGURE CAPTIONS

**Figure 1: Illustration of stabilizing and directional trait selection in ecological communities**.
The blue curves represent the ecological fitness of different species (mean reproductive success) as a function of their trait values and the bars indicate species trait distribution in the species pool (one value per species). Stabilizing selection is when intermediate trait values of the trait distribution of the species pool matches the peak of the fitness curve, while directional selection is when the highest fitness is found under extreme trait values (in this example, for the highest values). Selection strength represents the degree of fitness reduction with increasing distance from the optimal value, where a steeper decline indicates stronger selection because of higher interspecific fitness differences.

**Fig. 2**: A scheme for inferring selection attributes based on the comparison between observed patterns and null expectations. A) If species diversity is lower than the null expectation then selection has occurred (the larger the difference the stronger the selection). B) If community weighted variance (CWV) of a particular trait is lower than the null expectation it implies a trait-specific selection for that trait (the larger the difference the stronger the selection) C) If community weighted mean (CWM) of a particular trait differ from the null expectation it implies a directional selection. Alternatively, if CWM is similar, it implies a stabilizing selection (assuming the previous step has shown that CWV is lower than expected).

**Figure 3:** The community-level outcomes of varying selection attributes in the simulation model. (A) community weighted mean (CWM) of trait values diverge from the species pool mean (0) only under directional selection (B) community weighted variance (CWV) is affected

by selection strength. Species diversity is affected by both directionality and selection strength: (C) Inverse Simpson diversity at the local community scale (α); (D) Inverse Simpson diversity in the metacommunity scale (γ); (E) Species richness in the local community scale (α); (F) Species richness in the metacommunity scale (γ). Note the logarithmic scale of the x-axes (all panels) and some y-axes (panels C-F).

**Figure 4:** Ecological selection in the mesocosm experiment. Circles are values from each experimental community (n=18) while triangles represent the means of each soil depth treatment. Dashed lines represent the null expectations (a simulation envelope representing the extremum of 95% of the runs) (A) Species richness is lower in the deep soil treatment (productive habitat) compared with the shallow soil treatment (less productive) but lower than the null expectation in both habitat types. (B) Inverse Simpson's diversity is lower in the deep soil treatment, but in both habitat types, observed values are lower than the null expectations. (C) Community weighted mean seed mass [(log$_{10}$ mg)] is higher than the null expectation in the deep soil treatment. (D) Community weighted variance is higher than the null expectation in the shallow soil and lower than expected in the deep soil treatment. Note the logarithmic scale of the y-axes for diversity indices. Differences between the habitats are statistically significant ($P<0.05$ for all comparisons based on permutation tests using the R package 'Perm' for avoiding heteroscedasticity). See methods for details on the different null expectations for species diversity (based on a dynamics simulation of pure drift) and trait distributions (based on observed patterns of species distribution and the shuffling of trait values).

**Online supporting information for the paper 'the effects of ecological selection on species diversity and trait distribution: predictions and an empirical test'**

**Ecology**

Niv DeMalach, Po-Ju Ke & Tadashi Fukami

## Appendix S1 – Investigating the theoretical model

This appendix includes additional simulations of our model to gain a deeper understanding of its behavior and test its robustness to different assumptions. Specifically, we investigated: (a) temporal dynamics (Fig. S1-S3), (b) the effects of varying the underlying trait distribution Fig. S4-S6), (c) the effects of varying the fitness function (Fig. S7-S9), and (d) a scenario when the trait being measured is not the trait being selected (Fig. S10).

*(a) Temporal dynamics*

Here, we investigate temporal dynamics from the beginning until the end of the simulation (5000 timesteps) for understanding transient dynamics. We investigate these transient dynamics because a major challenge in testing theoretical models using experimental systems is that the timescales of experiments often represent transient dynamics rather than the equilibrium conditions. In our model, reaching an equilibrium could take up to 3000 timesteps under very low stabilizing selection (Fig. S1) but can be much faster under strong selection (Fig. S2). Additionally, we tested whether qualitatively similar results could be obtained in 10 generations (years). We found that the main results after a decade are qualitatively similar to the main results in equilibrium (Fig. S3), suggesting that difference in species diversity between stabilizing and directional selection could be detected within the timescales of the experiment when selection is strong enough.

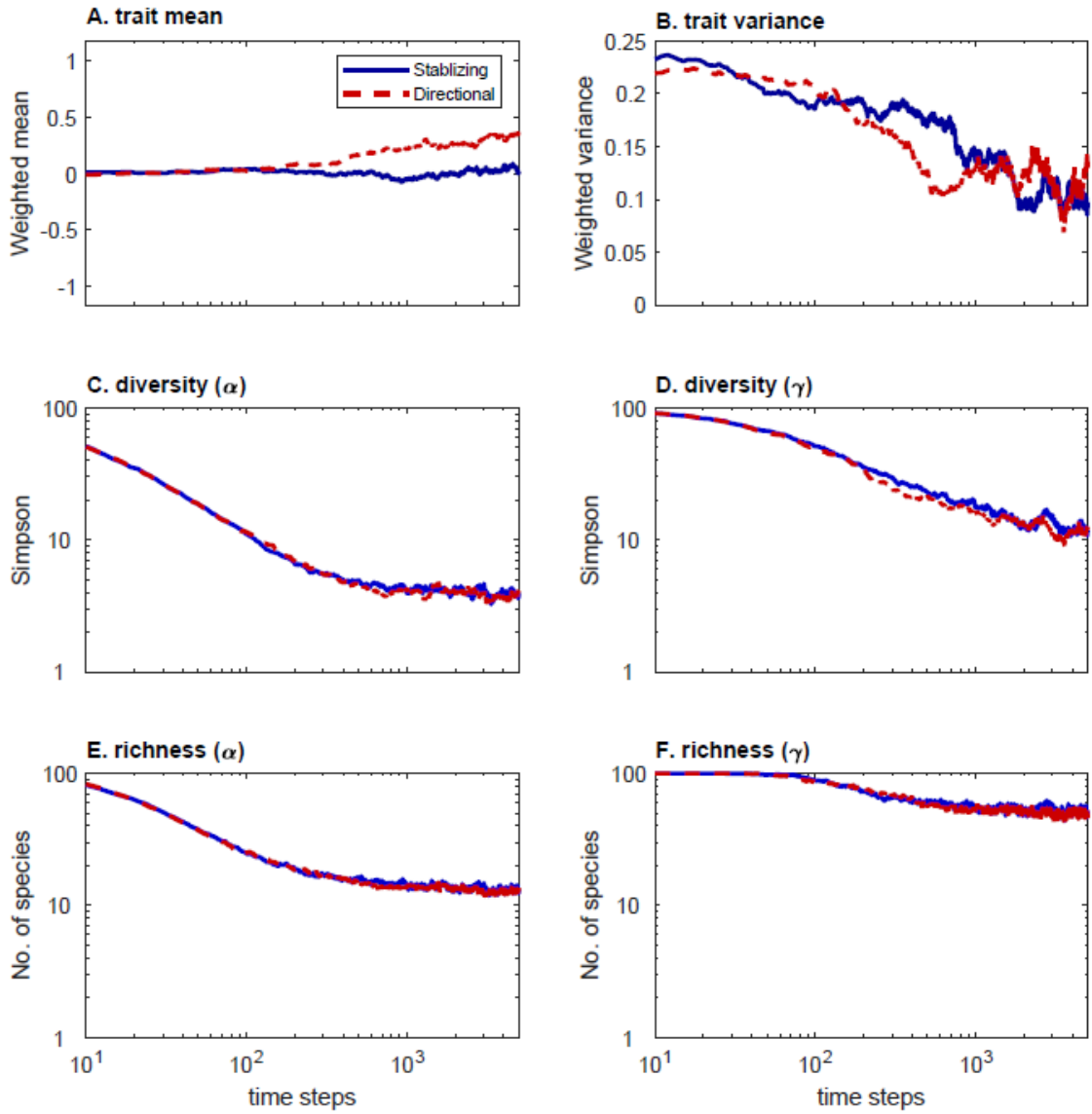

**Figure S1:** Community attributes along time under the lowest selection level ($\theta = 10^{-3}$). Note the logarithmic scales

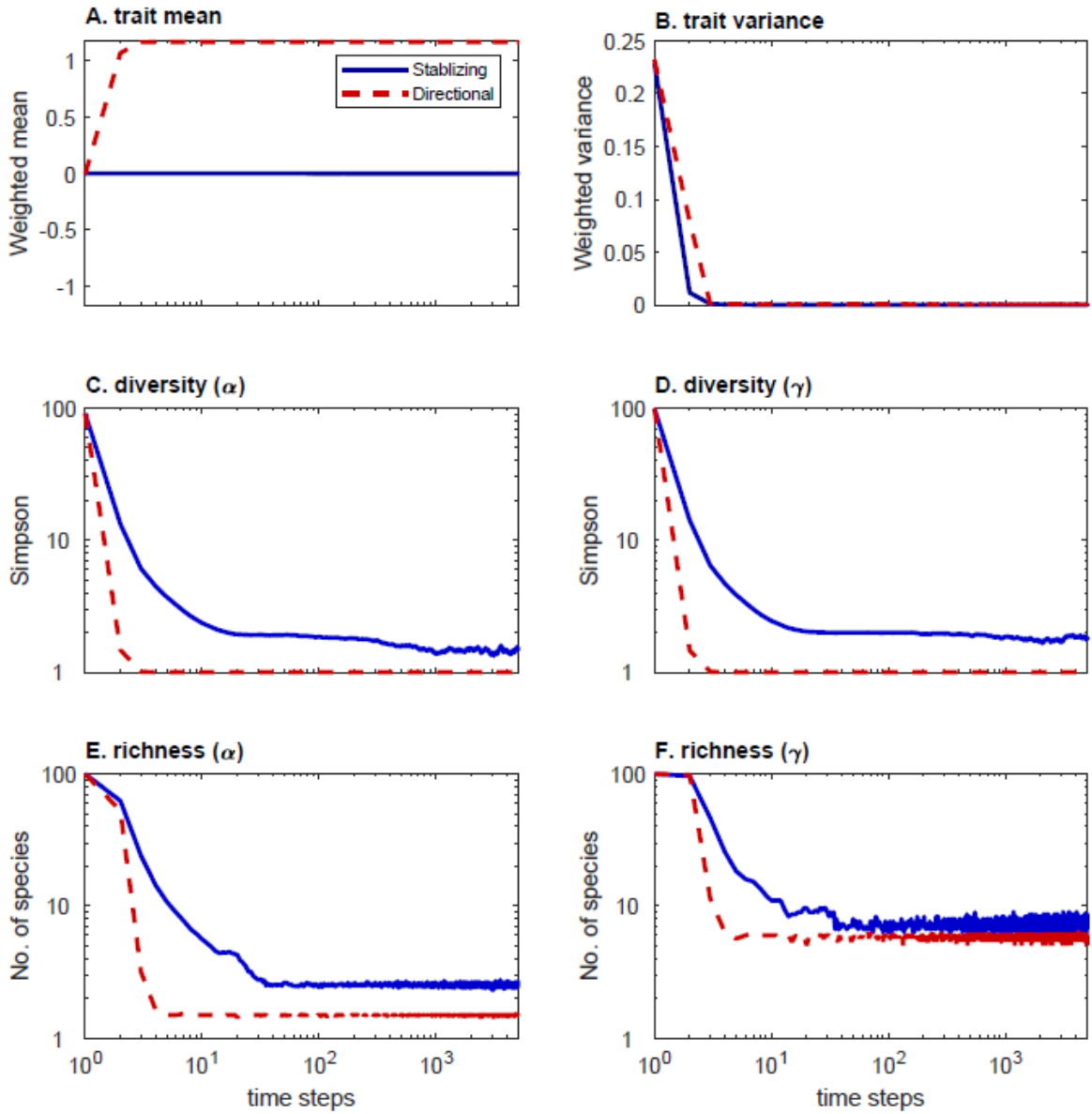

**Figure S2:** Community attributes along time under the highest selection level ($\theta = 10^3$). Note the logarithmic scales

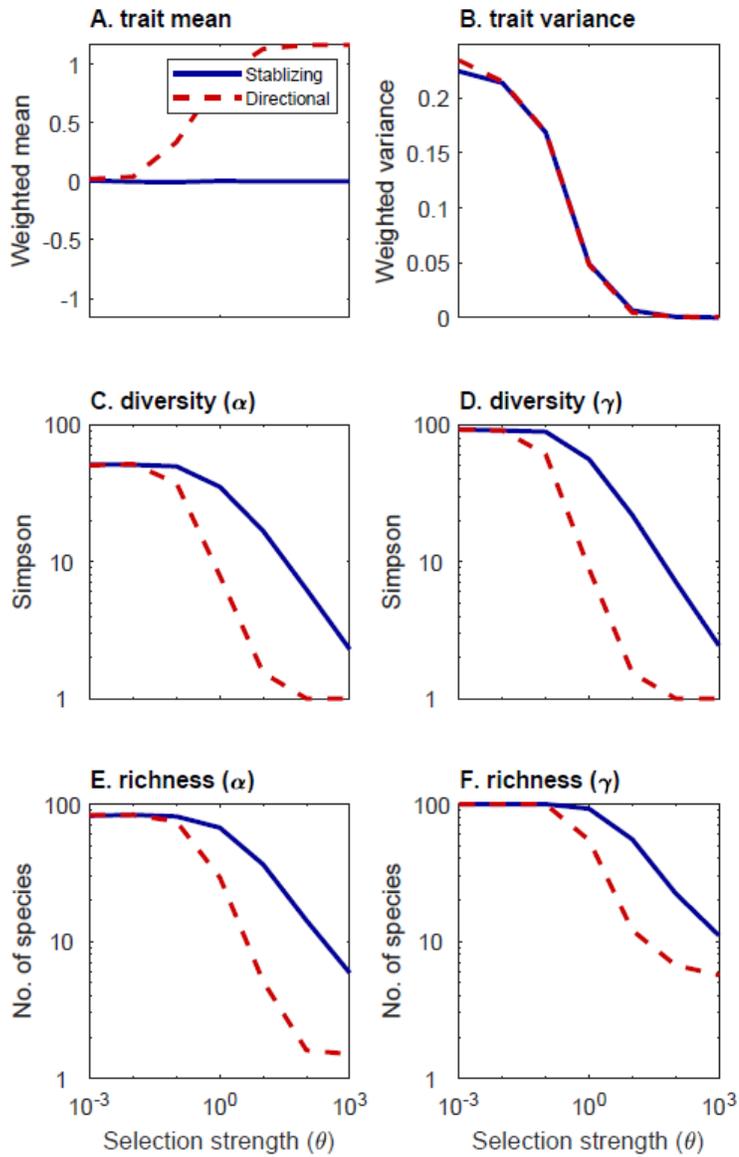

**Figure S3:** The effect of selection strength on community attributes under stabilizing selection (blue solid lines) and directional selection (indicated by red dashed lines) after ten generations (years).

*(b) Trait distribution*

Here, we investigated scenarios where the standard deviation of trait distribution of the species pool is lower ($\boldsymbol{\delta_{SD}} = \mathbf{0.1}$) or higher ($\boldsymbol{\delta_{SD}} = \mathbf{1}$) than in the main simulation (where $\boldsymbol{\delta_{SD}} = \mathbf{0.5}$). We also tested how does uniform trait distribution affect the results (with a similar standard deviation to the main simulation, range [-0.8 0.8]). While we are unaware of any empirical species pool with a uniform distribution, this simulation was used for inferring whether our predictions stem from the assumption of normal distribution.

We found that changing $\boldsymbol{\delta_{SD}}$ has no qualitative effect on community attributes but there are few quantitative effects (Fig. S4-5). First, higher levels of $\boldsymbol{\delta_{SD}}$ could lead to higher community weighted variance (CWV) but this effect decreases with increasing $\boldsymbol{\theta}$ (when selection is very strong, CWV is low regardless of $\boldsymbol{\delta_{SD}}$).

Another effect of increasing $\boldsymbol{\delta_{SD}}$ is decreasing species diversity because higher trait differences lead to higher fitness differences. Lastly, there are three-way interactions between selection strength ($\boldsymbol{\theta}$)**,** selection type (stabilizing vs. directional) and $\boldsymbol{\delta_{SD}}$. An inevitable outcome of equation (2) is that there are no differences between stabilizing and directional selection when $\boldsymbol{\theta} = \mathbf{0}$. However, the degree of selection strength where a difference between stabilizing and directional selection can be detected depends on $\boldsymbol{\delta_{SD}}$ . As $\boldsymbol{\delta_{SD}}$ increases, lower levels of $\boldsymbol{\theta}$ are needed for a difference between the two selection types to appear.

Patterns of trait distribution under a uniform distribution were very similar to the patterns obtained under a normal distribution (compare Fig. S6 with Fig. 2 in the main text). However, species diversity patterns differed, i.e. there were smaller (but still apparent) differences between stabilizing and directional selection under the uniform distribution. The finding that the differences did not disappear implies that differences in species diversity patterns between stabilizing and directional selection are not a mere result of the assumption of normal trait distribution (they are also related to geometric constraints).

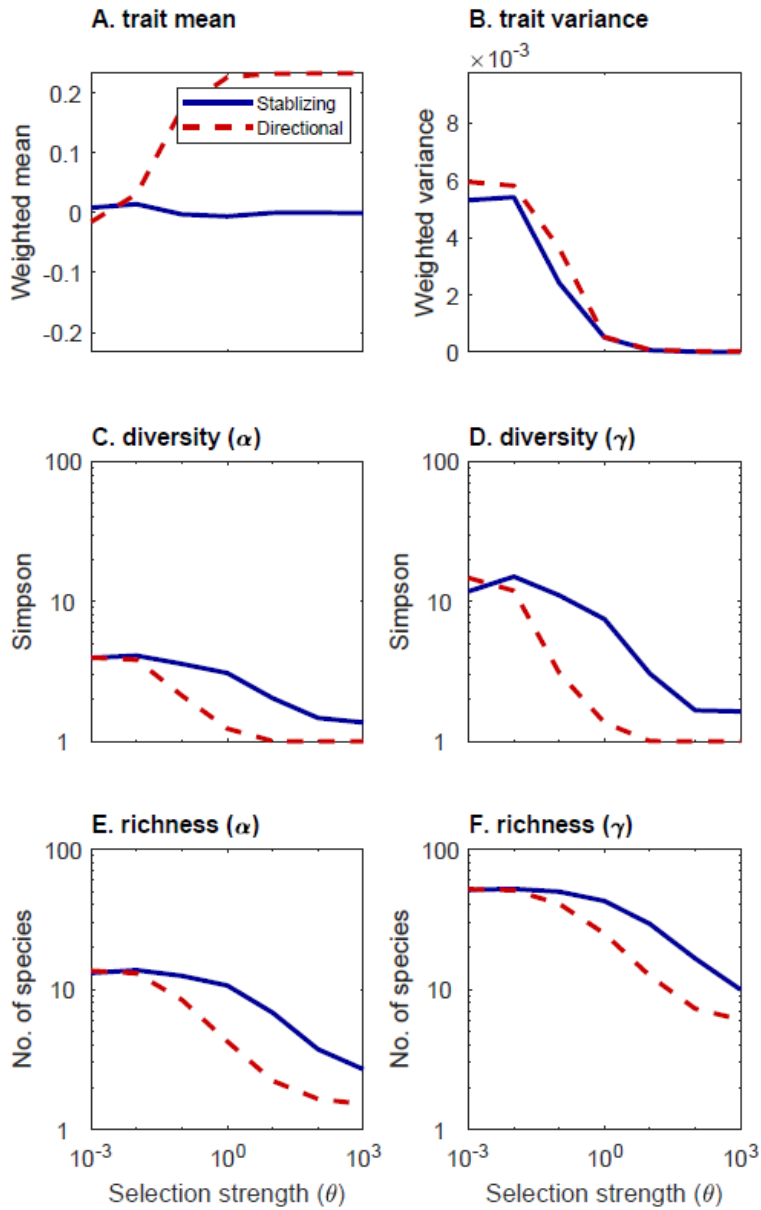

**Figure S4:** The effect of selection strength on community attributes under stabilizing selection (blue solid lines) and directional selection (by red dashed lines) when $\delta_{SD}= 0.1$. (lower trait variation in the species pool compared with the main simulation as shown in Fig. 2).

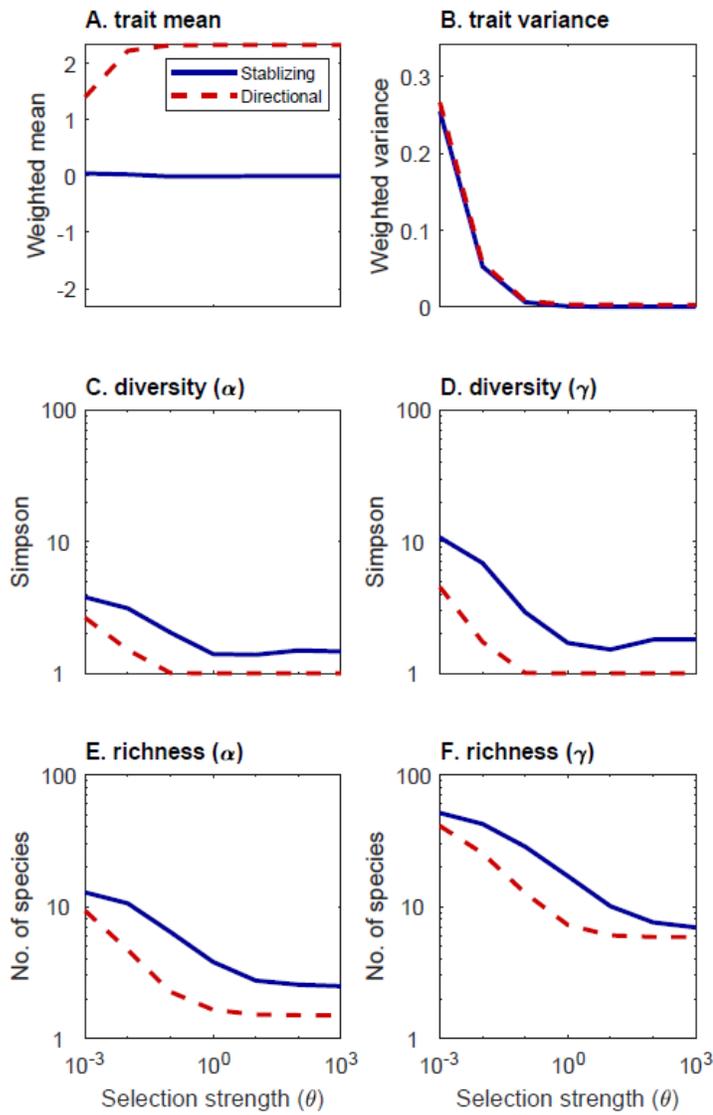

**Figure S5:** The effect of selection strength on community attributes under stabilizing selection (blue solid lines) and directional selection (red dashed lines) when $\delta_{SD}= 1$. (higher trait variation in the species pool compared with the main simulation as shown in Fig. 2)

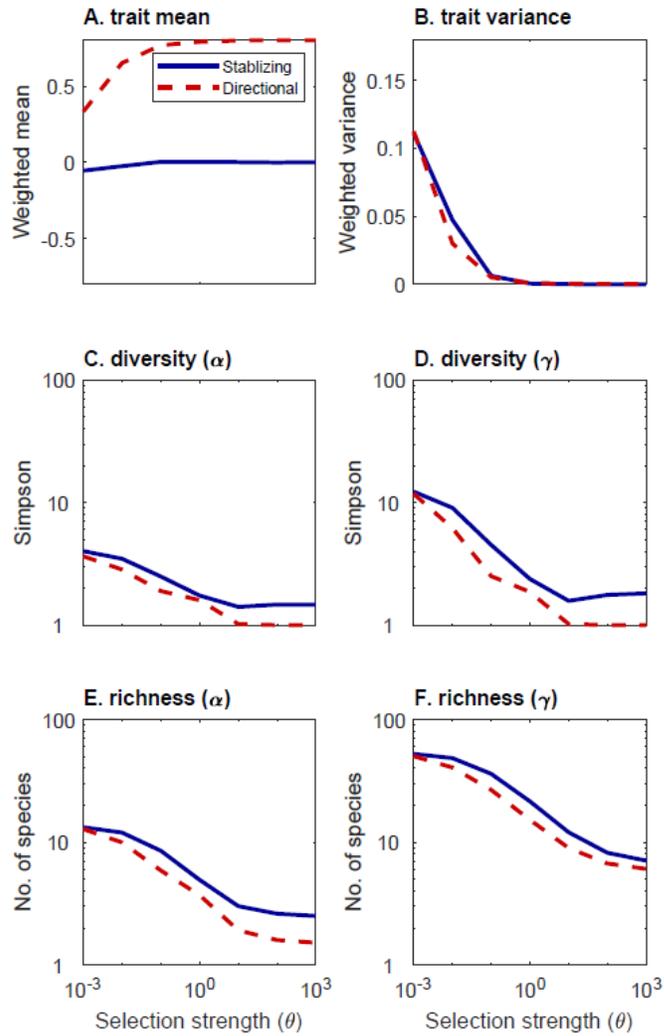

**Figure S6:** The effect of selection strength on community attributes under stabilizing selection (blue solid lines) and directional selection (red dashed lines) when trait distribution in the species pool is uniform.

*(c) the effects of a different fitness function*

We investigated the sensitivity of our modeling approach to the specific function used for relating traits and fitness by substituting equation (2) in the main text with the following equation:

$$\omega_i = \max[1 - \theta(\delta_{best} - \delta_i)^2, 0] \qquad ,$$

$$(3)$$

where $\delta_{best}$ represent the optimal trait value and $\theta$ determine the strength of the selection, i.e. the degree of the fitness differences for a given trait distance. As in equation (2), when $\theta = 0$ all species have equal fitness, and as $\theta$ increases, the larger the fitness differences. However, unlike the Lorentzian function used in the main text, which is naturally constrained between zero and one, the function used here includes a minimum term for avoiding negative values (it is more similar to selection functions being used in population genetics). This minimum term implies that instead of fitness asymptotically approaching zero with increasing distance from the optimum value (Fig. S7), fitness values are set to zero when trait distances are above some threshold distance (Fig. S8). Therefore, while the interpretation of $\theta$ as selection strength applies for both equations (2) and (3), the actual values are not comparable.

Changing the fitness function has no qualitative effect on the results. In both cases, species diversity is higher under stabilizing selection (Fig. S9). Besides, in both cases, CWV is similar for both types of selection. The only difference found following this change is that CWV is slightly higher under directional selection compared with stabilizing selection despite the latter leading to higher species diversity.

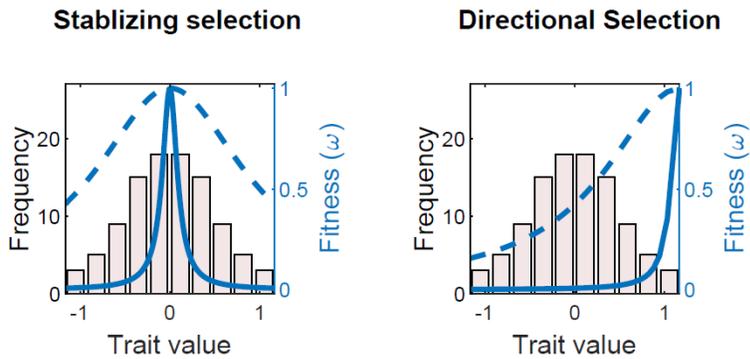

**Figure S7:** Illustration of stabilizing and directional selection types using the fitness function applied in the main text (equation 2). The dashed and the solid lines describe fitness ($\omega$) as a function of trait values under two levels of selection strength ($\theta = 1$ and $\theta = 100$, see equation 2 for details). The histograms in the background show the trait distribution ($\delta_{mean} = 0$, $\delta_{SD} = 0.5$) used in the simulation model. Note that left y-axes refer the histogram (frequency) and the right y-axes refer to the curve (fitness).

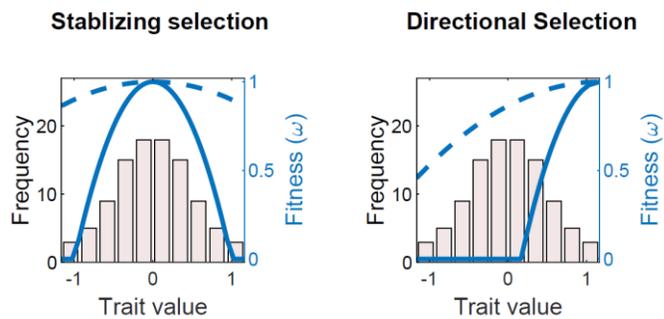

**Figure S8:** Illustration of stabilizing and directional selection types using alternative fitness function (equation 3). The dashed and the solid lines describe fitness ($\omega$) as a function of trait

values under two levels of selection strength ($\theta = 0.1$ and $\theta = 1$ , see equation 3 for details). The histograms in the background show the trait distribution ($\delta_{mean} = 0$, $\delta_{SD} = 0.5$) used in the simulation model. Note that left y-axes refer to the histogram (frequency) and the right y-axes refer to the curve (fitness).

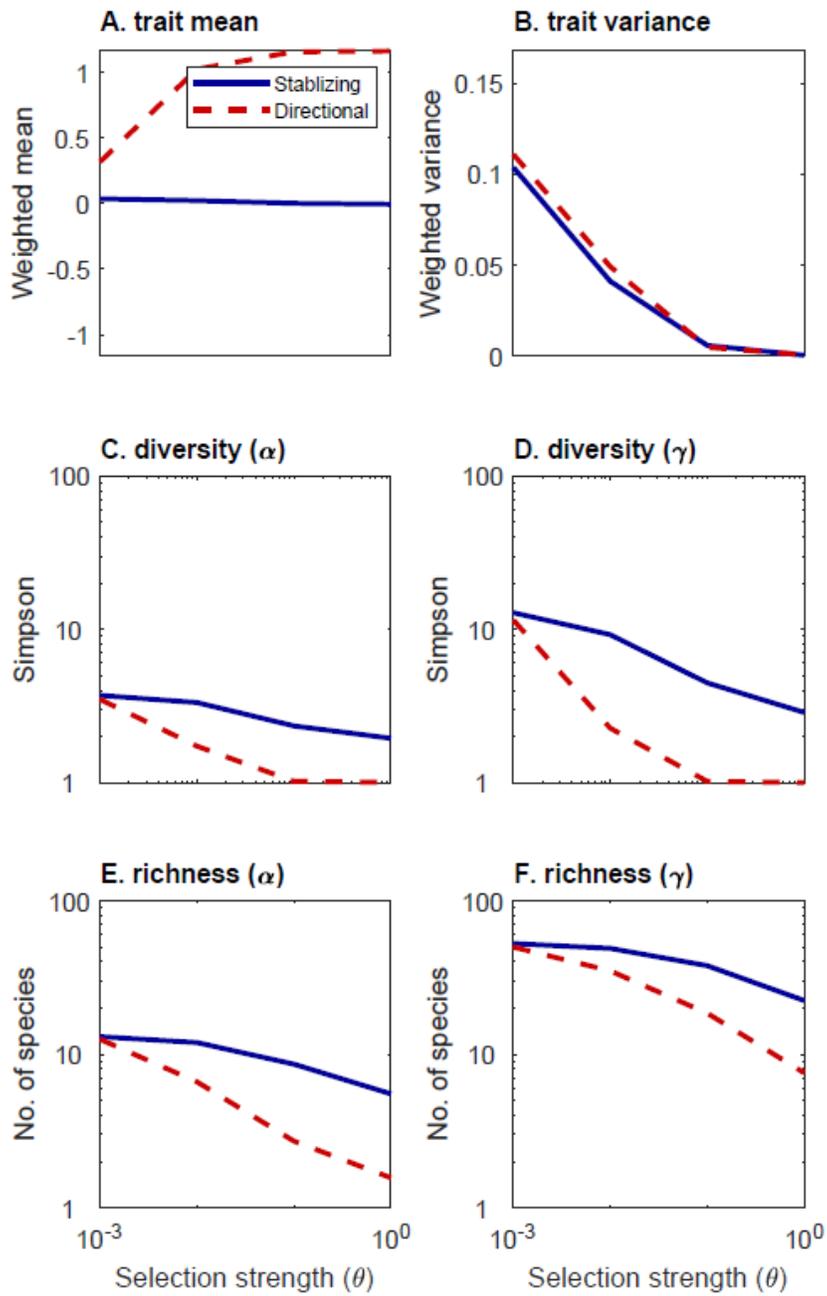

**Figure S9:** The effect of selection strength on community attributes under stabilizing selection (blue solid lines) and directional selection (red dashed lines) using alternative fitness function (equation 3).

*(d) wrong trait scenario*

Here, we investigated a situation where the main trait being investigated is not the trait being affected by selection (selection affects an unmeasured trait). For this aim, we randomly permutated the fitness values (based on equation 2) among the species thereby breaking the correlation between fitness and trait value. This approach leads to much larger differences among simulations (because by chance species with extremely small or high value can have high fitness). We found that under this scenario both types of selection do not affect the community mean while both selection types reduce the variance as selection strength increases, although the variance is lower under directional selection for any given level of selection strength (Fig. S10).

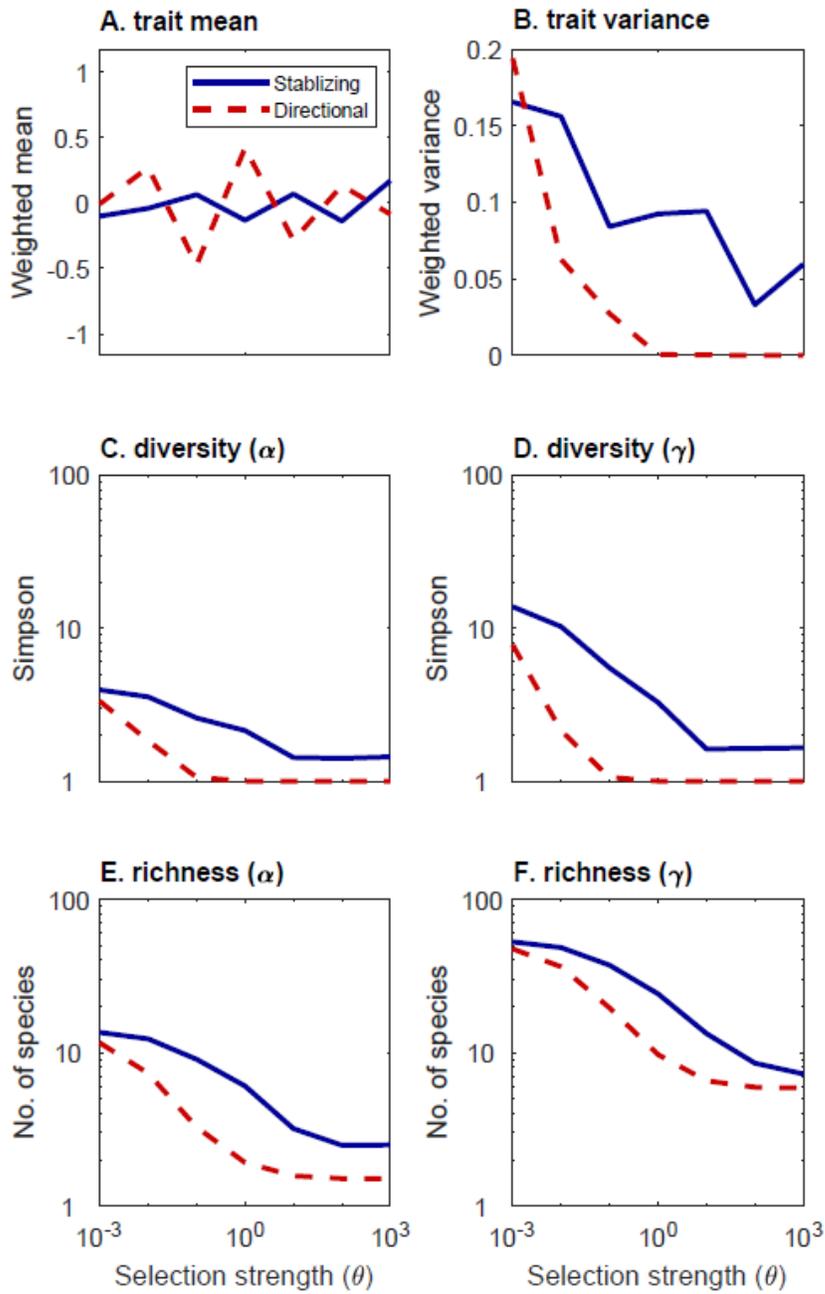

**Figure S10:** The effect of selection strength on community attributes when the trait being measured is independent of the trait being selected.

## Appendix S2: the mesocosm experiment

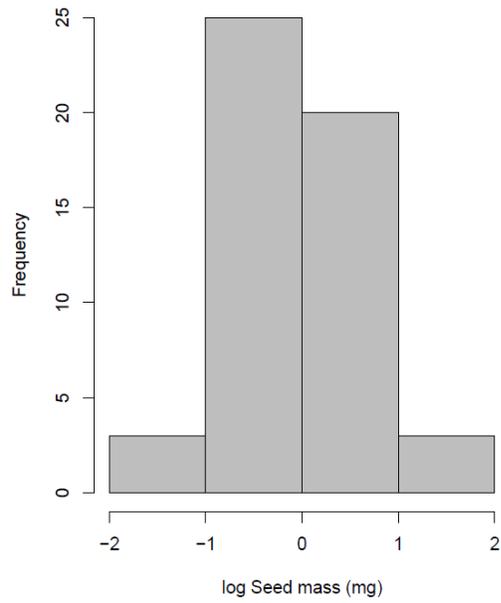

**Figure S1:** Seed mass distribution (in milligrams after log10 transformation) of the species sown in the mesocosm experiment.

**Table S1: List of the 51 species used in the mesocosm experiment.**

| Species name | Family | Seed Mass (mg) |
|---|---|---|
| *Alyssum minus* | Brassicaceae | 0.739 |
| *Anagallis arvensis* | Primulaceae | 0.484 |
| *Anthemis hebronica* | Asteraceae | 1.350 |
| *Astragalus hamosus* | Fabaceae | 2.943 |
| *Atractylis cancellata* | Asteraceae | 1.460 |
| *Avena sterilis* | Poaceae | 17.861 |
| *Calendula arvensis* | Asteraceae | 1.555 |
| *Catapodium rigidum* | Poaceae | 0.195 |
| *Chaetosciadium trichospermum* | Apiaceae | 1.091 |
| *Chrysanthemum coronarium* | Asteraceae | 2.471 |
| *Crepis aspera* | Asteraceae | 0.402 |
| *Crucianella macrostachya* | Rubiaceae | 1.346 |
| *Daucus broteri* | Apiaceae | 1.645 |
| *Delphinium peregrinum* | Ranunculaceae | 0.903 |
| *Filago contracta* | Asteraceae | 0.107 |
| *Filago pyramidata* | Asteraceae | 0.036 |
| *Galium setaceum* | Rubiaceae | 0.104 |
| *Geropogon hybridus* | Asteraceae | 27.234 |
| *Hedypnois rhagadioloides* | Asteraceae | 1.434 |
| *Helianthemum ledifolium* | Cistaceae | 0.392 |
| *Helianthemum salicifolium* | Cistaceae | 0.145 |
| *Hirschfeldia incana* | Brassicaceae | 0.293 |
| *Hordeum spontaneum* | Poaceae | 26.041 |
| *Hymenocarpos circinnatus* | Fabaceae | 6.656 |
| *Lagoecia cuminoides* | Apiaceae | 0.498 |
| *Linum strictum* | Linaceae | 0.249 |
| *Lolium rigidum* | Poaceae | 2.147 |
| *Lomelosia prolifera* | Dipsacaceae | 5.244 |
| *Lotus peregrinus* | Fabaceae | 2.655 |
| *Medicago (trigonela) monspeliaca* | Fabaceae | 0.740 |
| *Medicago orbicularis* | Fabaceae | 5.246 |
| *Minuartia decipiens* | caryophyllaceae | 0.250 |
| *Misopates orontium* | Scrophulariaceae | 0.120 |
| *Papaveraceaer umbonatum* | Papaveraceae | 0.099 |
| *Picris longirostris* | Asteraceae | 1.395 |
| *Pimpinella cretica* | Apiaceae | 0.265 |
| *Plantaginaceaeo afra* | Plantaginaceae | 0.799 |
| *Plantaginaceaeo cretica* | Plantaginaceae | 1.125 |
| *Plantaginaceaeo lagopus* | Plantaginaceae | 0.417 |
| *Pterocephalus brevis* | Dipsacaceae | 0.803 |
| *Rhagadiolus stellatus* | Asteraceae | 1.388 |
| *Sedum rubens* | Crassulaceae | 0.066 |
| *Stachys neurocalycina* | Lamiaceae | 0.999 |
| *Stipa capensis* | Poaceae | 1.410 |
| *Telmissa microcarpa* | Crassulaceae | 0.288 |
| *Tordylium trachycarpum* | Apiaceae | 0.200 |
| *Torilis tenella* | Apiaceae | 0.607 |
| *Trifolium purpureum* | Fabaceae | 1.307 |
| *Trifolium tomentosum* | Fabaceae | 0.556 |
| *Urospermum picroides* | Asteraceae | 1.718 |
| *Valantia hispida* | Rubiaceae | 0.394 |

Table S2: parameters of the drift model for the mesocosm experiment

| Symbol | Description | Value(s) |
|---|---|---|
| $S$ | Species pool size (number of species) | 47 |
| $N$ | Local community size (number of individuals) | 1661 |
| $n$ | Number of local communities | 9 |
| $Timesteps$ | Length of the simulation (generations) | 5 |
| $\delta_{mean}$ | Trait mean ($\log_{10}$ seed mass) | -0.10 |
| $\delta_{SD}$ | Standard deviation trait ($\log_{10}$ seed mass) | 0.62 |
| $\theta$ | Selection strength (dimensionless) | 0 |
| $D$ | The proportion of dispersed individuals | 0 |
| $I$ | Proportion of immigrants | 0 |